\documentclass[5p,authoryear]{elsarticle}
\usepackage{amsmath}
\usepackage{amssymb}
\usepackage{eurosym}
\usepackage{color}
\usepackage{hyperref}
\numberwithin{equation}{section}
\usepackage{subcaption}
\usepackage{graphicx}

\begin{document}
\author[hevs]{Marc Gillioz\corref{cor}}
\author[hevs]{Guillaume Dubuis}
\author[arma]{\'Etienne Voutaz}
\author[hevs,dqmp,andlinger]{Philippe Jacquod}
\address[hevs]{School of Engineering, University of Applied Sciences of Western Switzerland HES-SO, 1951 Sion, Switzerland}
\address[arma]{Cyber-Defence Campus, armasuisse, Feuerwerkerstrasse 39, 3604 Thun, Switzerland}
\address[dqmp]{Department of Quantum Matter Physics, University of Geneva, 1211 Geneva, Switzerland}
\address[andlinger]{Andlinger Center for Energy and the Environment, Princeton University, Princeton, NJ 08544 USA}

\cortext[cor]{Corresponding author: marc.gillioz@hevs.ch}

\title{Anomaly Detection with Machine Learning Algorithms in Large-Scale Power Grids}
\begin{abstract}
We apply several machine learning algorithms to the problem of anomaly detection in operational data for large-scale, high-voltage electric power grids. 
We observe important differences in the performance of the algorithms.
Neural networks typically outperform classical algorithms such as k-nearest neighbors and support vector machines, which we explain by the strong contextual nature of the anomalies.
We show that unsupervised learning algorithm work remarkably well and that their predictions are robust against simultaneous, concurring anomalies.

\end{abstract}
\date{\today}

\maketitle

\section{Introduction} 
\label{sec:introduction}

Electric power grids are undergoing fundamental changes in the way they operate. Decarbonization of the energy system requires a transition to renewable 
generation of electric power, which  replaces centralized and dispatchable generators based on  grid-stabilizing rotating machines, with intermittent, geographically distributed 
and inertialess generation~\citep{CO2Emissions2022}. Guaranteeing the safety of supply of electric power and the stability of present and future power grids  poses a number of challenges~\citep{Anees2012, Milano2018, Basit2020, Denholm2020, Makolo2021, Kerci2023}. In particular, 
transitioning power grids operate more often closer to their operational limit~\citep{Martinez2025}, which requires fast and adequate remedial actions from the part of the operator. To enable fast reaction times and efficient remedial actions, power grids are further modernized with smart grid technologies, including advanced sensors and enhanced communication systems for real-time monitoring and control~\citep{Ohanu2024}. 

Real-time monitoring and control requires that the operator knows the state of the system as accurately as possible -- they need to receive large, multivariate sets of data, e.g.~on power injection, voltage amplitude, phase and frequency at each substation, on the power flow on each line and so forth. The reliability of these data needs to be guaranteed. Simultaneously, 
rare events where some of these data deviate significantly from their usual range need to be detected as fast as possible, in order to guarantee a prompt, appropriate response of the operator, that would not jeopardize grid stability. There is thus an urgent need for fast algorithms for efficient data anomaly detection~\citep{Thudumu2020}. Data on the operational state of the system can be corrupted and anomalies may occur because of malfunctioning sensors or communication systems, or following a deception cyber-attack
with false data injection~\citep{Pasqualetti13, Liu2012}. Such occurrences will most certainly multiply in the future, because the massive deployment of smart grid technologies multiply the number of entrance doors to the communication system and the probability that one or few electronic component malfunctions~\citep{Ogie2017, IEAcyber}. Other types of anomalies may reflect rare events following unplanned power line outages or the sudden disconnection of a generator, or simply occur due to erroneous data acquisition or transmission. Being purely data-based and fast, machine learning (ML) algorithms are natural candidates for anomaly detection, regardless of the origin of the anomaly~\citep{PIMENTEL2014}.
In this article, we compare the performances of nine different ML algorithms to detect anomalies in multivariate time series corresponding to 
power injection into large, high-voltage transmission power grids. These investigations are based on earlier works, where we constructed a model 
for the synchronous transmission power grid of continental Europe~\citep{PanTaGruEl, Pagnier2019, Tyloo2019} and where
we developed a method for generating large sets of statistically realistic data for such power grids, intended to be used for training and testing of 
ML algorithms~\citep{zenodo, Gillioz2025}. These tools allow us to present extensive tests of these different algorithms, spanning a large range of
different supervised and unsupervised approaches, and compare their performance for anomaly detection. 

\newpage

ML approaches have already been applied to electric power grids and here we mention a few works.
A number of investigations constructed and trained ML algorithms for evaluating the safety and reliability of operation via state estimation~\citep{Arteaga2019, Fioretto2020, Misyris2020, Duchesne2020, Stiasny2022, Guddanti2023, Hamann2024, Varbella2024}. While not directly related to the present work, these investigations are important, because their success demonstrates the ability of well-trained ML algorithm to reproduce true power flow data, which is a clear prerequisite for anomaly detection. 
In the context of electric power grids, the task of anomaly detection is to identify 
abnormal data in multiple time series, e.g.~corresponding to power injection at all buses on a given power grid. 
Data can be anomalous because of 
(i) faulty data acquisition or transmission, (ii) a deception cyber-attack, or (iii) anomalous operation, where e.g.~a power plant is suddenly disconnected from the grid or produces 
more than its rated capacity. In all cases, it is important to differentiate absolute from 
contextual anomalies. Absolute anomalies correspond to data that significantly differ from normal behaviors - for instance, because they lie  several standard deviation away from the multidimensional distribution of data, or in a lower density region of the distribution of normal data. They 
are evidently easy to detect.
On the other hand, a vanishing power injection from a plant at a given time is a contextual anomaly. It is not anomalous {\it per se}, but  only with respect to the general context, i.e.~the landscape of power injections from all power plants and of power consumptions from all consumer buses. 
The question arises as to whether contextual anomalies in power grids 
can  be detected solely from the simultaneous data pertaining to the rest of the grid.

There have been many works on anomaly detections in power grids~\citep{Zhang2021}, using methods ranging from classical time series analysis~\citep{Chou2014}, to
statistical approaches~\citep{Wei2020} and ML approaches. The general strategy is to construct a forecasted distribution of normal data and identify data points lying on lower density regions or even outside of that distribution. The principal difficulty arises from the multidimensionality of the data time series, with intricate cross-correlations 
between different time sub-series, which govern their contextual behavior but are hard to quantify. It is in particular for that reason that ML approaches have been intensively
pursued. We give here an inevitably incomplete survey of existing works.  Different ML algorithms have been successful for anomaly detections.
A support vector machine algorithm has been applied in \cite{Esmalifalak17} to detect
anomalies arising from deception attacks. The performance was quantified by $F_1$-scores reaching values $\approx 0.83$ and exceeding classical statistical detections methods for long enough training on the IEEE 188-bus test system. 
An unsupervised isolated forest algorithm was used in \cite{Mao2018}, which showed encouraging results with an accuracy $\gtrsim 0.7$
with relatively short training sequences on direct power consumption data for 160 users.
Significantly better performances were obtained with a random forest algorithm, reaching accuracy and detection rate above 90~\% in \cite{Wang2019}.
Further works applied recurrent and deep neural networks to anomaly detection, with performances also well above 90~\% for precision, recall and
$F_1$-score~\citep{Wilson2018, Gopali2021, Ding2021}. \cite{Dai2022} constructed an unsupervised method based on deep generative models augmented by 
Bayesian networks encoding causal relationships among the constituent time series. The method was compared to existing 
deep methods~\citep{Malhotra2016, Ruff2018, Sabokrou2020, Goyal2020, Ruff2020} and shown to have better performances, with ROC AUC scores reaching 67 to 80~\%
depending on the considered datasets.
Machine learning algorithms have also been applied to other tasks, such as the detection, localization, and prevention of short circuits \citep{Rafique2021, Belagoune2021, Gjorgiev2023}.

Most of these works focused on either small systems or distribution grids.
Most also considered absolute anomalies. In this manuscript we present investigations on three different, large transmission grids where we compare the performances of  
nine different ML algorithms to detect contextual anomalies on power generation.  
We examine different types of input data, including time series for loads and production at all buses, but without redundant information such as power flows on the lines. We consider anomalies that are instantaneous -- they occur at a given time step following a period of regular operation -- and maximal within the rated power range. Such anomalies could represent power grid malfunctions or false data injection attacks with the goal of destabilizing the network. Two of our algorithms are unsupervised and therefore also sensitive to other types of anomalies.
The focus is primarily on hydroelectric power plants, as their ability to follow the load generally leads to more erratic production profiles that are particularly difficult to predict, however we consider other production sources as well.

The paper is organized as follows. Section~\ref{sec:methodology} describes the methodology, including the dataset that serves as ground truth, the definition of anomalies, and the presentation of the different algorithms. Our results are reported in Section~\ref{sec:results}, with emphasis on the performance of the algorithms depending on the choice of hyperparameters and input vectors.
An interpretation of the results is given in Section~\ref{sec:discussion}, together with general lessons for anomaly detection in large-scale power grids.

\section{Methodology}
\label{sec:methodology}

We present here our methodology. This begins with a description of the dataset and the different choices of input vectors. We proceed with the description of anomalies, either in the form of labeled attacks that are classified and tackled in a supervised setup, or of generic anomalies, using unsupervised time series forecasting.
Finally, we describe the different algorithms used for each task and specify the relevant range of hyperparameters.

\subsection{Input data}
\label{sec:input}

\begin{figure*}[t]
	\begin{subfigure}{0.4\linewidth}
		\includegraphics[width=\linewidth]{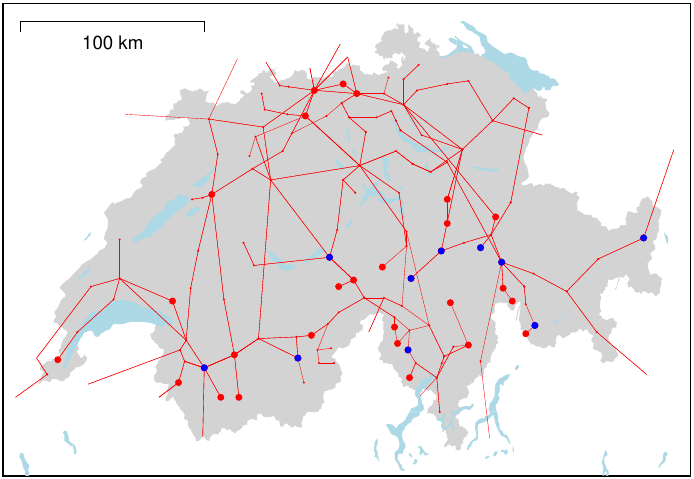}
		\caption{}
		\label{fig:network:CH}
	\end{subfigure}
	\hfill
	\begin{subfigure}{0.343\linewidth}
		\includegraphics[width=\linewidth]{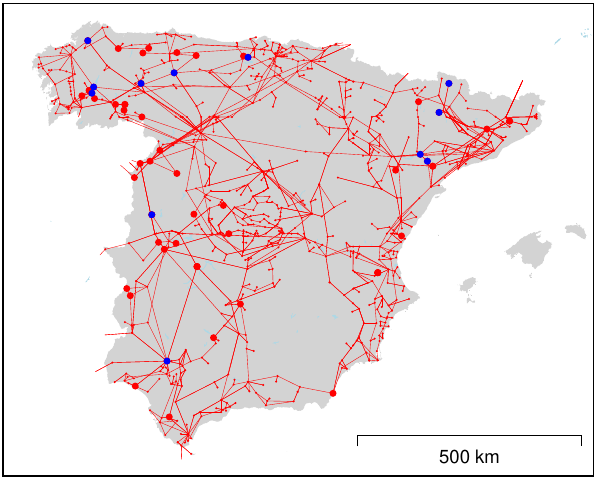}
		\caption{}
		\label{fig:network:ES}
	\end{subfigure}
	\hfill
	\begin{subfigure}{0.207\linewidth}
		\includegraphics[width=\linewidth]{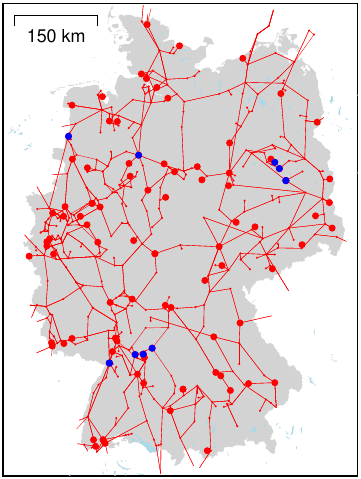}
		\caption{}
		\label{fig:network:DE}
	\end{subfigure}
	\caption{The three transmission power grids considered in this work:
	(a) Switzerland, (b) Spain, and (c) Germany.
	The dots represent power plants directly connected to the transmission grid,
	among which the blue ones are those on which we study anomalies.}
	\label{fig:networks}
\end{figure*}

We use as input the open access dataset \cite{zenodo}, which is described in \cite{Gillioz2025}. It consists of time series for production and consumption of active power at all the nodes of a large model of the transmission grid of continental Europe \citep{Pagnier2019, Tyloo2019}.
We focus on three subgrids corresponding to the national power transmission grids of
\begin{itemize}
	\item Switzerland, with 163 load buses and 36 generators; 
	
	\item Spain, with 908 load buses and 61 generators;
	
	\item Germany, with 560 load buses and 101 generators.
\end{itemize}
The topology of all three grids is shown in Figure~\ref{fig:networks}. We selected these three grids as test cases because they differ significantly in their loads,
their size, and their mix of power generation.
The dataset contains 20 years of data with hourly resolution. Variable loads are attached to all buses. The total load of each country matches the historical average, including daily, weakly, and seasonal modulations, but each individual load series is unique. This ensures that there are no spurious correlation between different loads.
The time series for production result from an optimal power flow computation with constraints on the rated power and total annual production of each power plant, and an objective function that balances the power flow through the lines in a realistic manner.
For highly-dispatchable power sources such as hydroelectric plants, this results in time series that appear erratic on short time scales, such as the one shown in Figure~\ref{fig:series}.

The large networks under consideration form the context of our work. We train and test our algorithms on a subset of all the available power plants. Since anomalies are obviously harder to detect and thus more interesting in highly-dispatchable sources, our main focus is on hydroelectric plants, although we also consider gas- and coal-fired power plants in Spain and Germany. We do not attempt to detect anomalies on nuclear power plants, which have mostly constant productions and are more closely monitored than other types of plants.
With its power generation originating at 50\% to 60\% from the hydroelectric sector, Switzerland is a good test case. We select for the study of anomalies ten hydroelectric plants, with rated power between 100 and 300~MW.
On the Spanish transmission grid, we select nine hydro plants with up to 400~MW of rated power, as well as one gas and two coal generators for comparison. On the German grid, we select 6 gas-fired power plants between 100 and 500~MW of rated power, as well as 5 coal-power plants with up to 1.5 GW of rated power.
The selected power plants in all three grids are located by blue dots on the three grid maps of Figure~\ref{fig:networks}.
It should be noted that only power sources directly connected to the transmission grid are taken into account. The increasing penetration of distributed renewable energy sources plays a minor role in our investigations, in that it lowers and modulates power consumption at load buses.

Our machine learning algorithms are trained once for each of the 33 selected power plants. We consider different type of inputs:
\begin{itemize}
	\item \emph{Context}: We consider both the restricted context
	consisting of all power generations only, and the more general context
	including all power injections (production and loads).
	Note that in the latter case the power flows through the national borders remain unknown,
	so that the anomalies cannot be detected by power balance arguments only.
	 
    \item \emph{History length}: Our input vectors range between a single time step
    of the grid at a given time $t$ (no history at all)
    to including several time steps, typically 4 hours and 24 hours.
    We also explore a wider historical time range with few selected algorithms,
    to determine whether anomaly detection can be improved by considering the grid
    history (see details in the discussion of Fig.~\ref{fig:history}).
    The built-in periodicity of the dataset means that we can always
    find a consistent history, no matter how long it needs to be.
    
    \item \emph{Historical context}: When taking history into account, 
    we always refer to the power plant under consideration. But we also explore 
    the possibility of  using the history of all production sources,
    or even of all grid injections.
\end{itemize}
We are interested in all combinations of input vectors proposed above. However, the size of the vectors increase rapidly when taking a large context (e.g.~968 entries for Spain with all injections) or when the entire historical context is included, and so does the computational time. Since we observe that the accuracy of anomaly detection decreases, as anticipated, when too much information is provided in the input to the algorithms, we do not study all cases extensively: for the Swiss network, we only consider the entire historical context up to 4 time steps of history and only with power generations as input;
for the Spanish and German grids, we do not take the entire historical context into account at all.
In spite of these restrictions, we consider sufficiently many different inputs to draw general conclusions.

\subsection{Supervised approach: false data injection attacks}
\label{sec:supervised}

The first anomalous scenario that we consider is that of a false data injection attack, see \cite{NESCOR2015}.
We assume that an attacker takes the control of the communication channels of a given power plant and transmits erroneous information to the grid operator. In the simplest case, the attacker can only modify the apparent power out of a single power plant. 
We focus on situations where the reported active power is altered at a given time $t$, with the time series at previous step unaffected. The action that maximizes the difference in power injection while respecting the known rated power of the plant is to pretend that the production is at full capacity if it is in reality operating at less than half its rated power, and conversely pretend that the plant is not producing when it is generating more than half its rated power.
Such ``on/off anomalies'' are optimal from the point of view of an attacker who does not have complete knowledge of the grid's state, in the sense that the largest power generation differences are expected to generate larger differences in the system, in particular in power flows on transmission lines.
This gives in turn the higher probability to make the operator believe that the grid has a problem.

\begin{figure*}[t]
	\begin{subfigure}{\linewidth}
		\includegraphics[width=\linewidth]{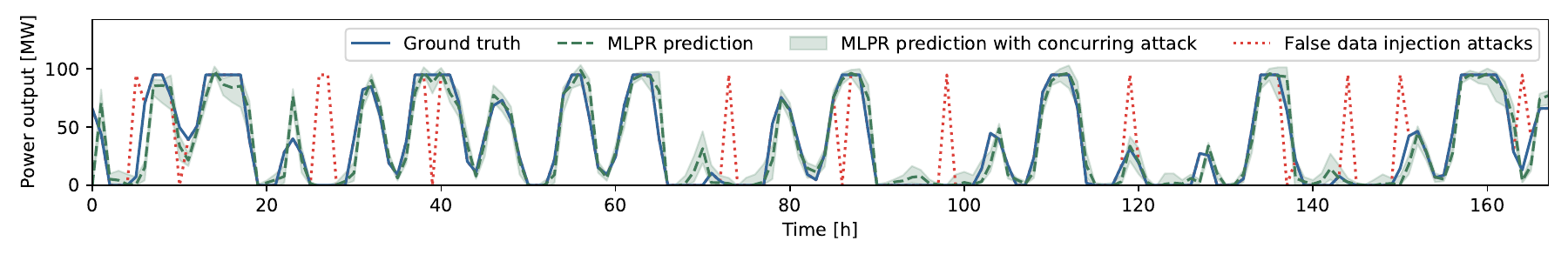}
		\caption{Löbbia hydroelectric power plant, Switzerland (2017, week 11)}
		\label{fig:series:Loebbia}
	\end{subfigure}
	\begin{subfigure}{\linewidth}
		\includegraphics[width=\linewidth]{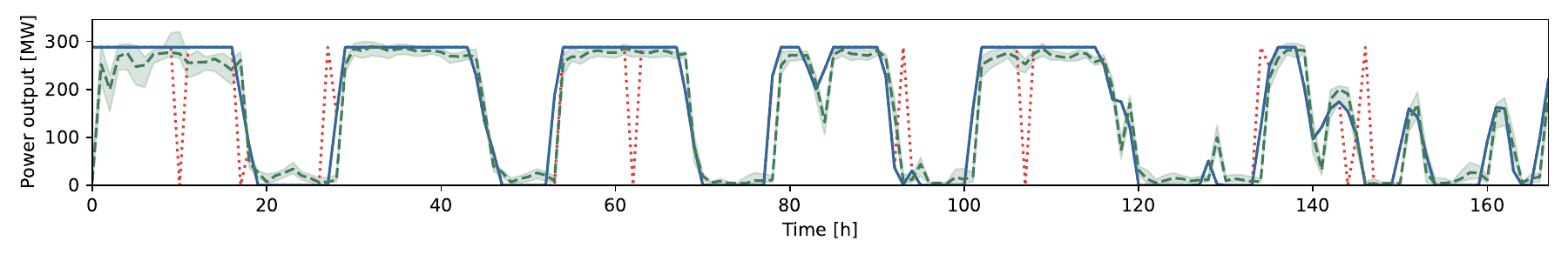}
		\caption{Pradella hydroelectric power plant, Switzerland (2017, week 8)}
		\label{fig:series:Pradella}
	\end{subfigure}
	\caption{Examples of synthetic time series for two hydroelectric power plant,
	as taken from the dataset~\citep{zenodo}.
	On/off anomalies added to the data are shown with a red, dotted line.
	The prediction of the unsupervised MLPR algorithm is shown
	with the green, dashed line.
	The green band surrounding the prediction indicates how much it can 
	deviate if a concurring attack happens on another generator
	in the same country.}
	\label{fig:series}
\end{figure*}

Supervised training data is therefore defined by altering 10\% of the time steps in our dataset following this rule and labeling those as anomalies. An example of resulting series is shown in Fig.~\ref{fig:series}.
Our algorithms are then trained to perform a simple binary classification into regular and anomalous categories.

The quality of the classification is measured with the $F_2$-score. Given that it is crucial from the point of view of the operator to detect every attack, this metric gives a higher weight to recall (rate of detecting actual attacks) than to precision (rate of false alarms). In terms of number of false positives FP (an alarm is raised in a regular situation), false negatives FN (no alarm is raised even though the situation is anomalous), and true positives TN, the $F_2$-score is defined as
\begin{equation}
	F_2 = \frac{5 \text{TP}}{5 \text{TP} + 4 \text{FN} + \text{FP}}.
	\label{eq:F2}
\end{equation}

Even though the supervised training stage focuses on false data injections at a given time $t$ and for the given power plant under scrutiny, with regular values both for the history and context, it is eventually possible to perform inference in the situation of multiple simultaneous attacks, where both the selected power plant and its context are altered. 

\subsection{Unsupervised approach: anomalies}
\label{sec:unsupervised}

In addition to the classification problem defined with "on/off" attacks, we also consider the more agnostic problem of predicting the value of a given production output  at time $t$ based on the history at previous time steps and on the context at the same time $t$.
This problem is more difficult, but the algorithms trained in this way are particularly valuable since they apply to all types of anomaly, and therefore well-suited for a broader range of applications.

In this case we use a regression based on the coefficient of determination $R^2$ to train the models. In order to compare their performance with the classifier models, we define in each case a detection threshold that is optimal with respect to the "on/off" attacks described in the previous sections. The threshold optimized in this way is quite large since the attacks are pronounced, typically a sizable fraction of the rated power. Lower thresholds based on the root mean square error of the models could be used to detect smaller anomalies. Here we have optimized our thresholds for anomalies of size of at least 50~MW.

\subsection{Algorithms and hyperparameters}
\label{sec:algos}

The classification and regression algorithms are implemented using standard libraries in Python. 
The first five algorithms rely on the \emph{scikit-learn} library \citep{scikitlearn}:

\begin{itemize}
	\item \textbf{NBC} --- The Gaussian Naive Bayes classifier is a simple method
	relying on the assumption that the likelihood of the features is Gaussian.
	It is not expected to perform well on the task at hand,
	but can be considered a baseline model for comparison with better algorithms.
	
	\item \textbf{KNNC} --- The $k$-nearest neighbors classifier
	is a non-generalizing algorithm relying on proximity with training data.
	The number of relevant neighbors $k$ is a hyperparameter that we vary between
	1 and 500 in exponentially growing steps.
	
	\item \textbf{SVC} --- The support vector machine classifier is a relatively 
	simple algorithm known to be effective in high-dimensional datasets.
	The regularization parameter $C$ is varied between 300 and 30,000
	in exponentially growing steps.
	
	\item \textbf{RFC} --- A random forest classifier with a number of decision
	trees that we vary between 20 and 100.
	
	\item \textbf{GBC} --- The gradient-boosted decision trees algorithm
	generalizes and often outperforms the random forest classifier.
	The number of boosting stages is taken in the range 10 to 1000.
	
	\item \textbf{MLPC} --- The multi-layer perceptron is a shallow neural network,
	with up to 4 hidden layers and up to 1000 neurons per layer.
\end{itemize}
This list is complemented with a deep learning algorithm implemented with the \emph{PyTorch} library \citep{pytorch}:
\begin{itemize}
	\item \textbf{LSTMC} --- This is a deeper neural network, particularly
	well-suited to handle time series, that consists of up to 3
	long short-term memory (LSTM) modules stacked on top of each other,
	each with 256 hidden features. We use a batch size of 128 and
	a dropout rate of 75~\% to prevent overfitting and improve the generalization
	capability.	
\end{itemize}
Finally, two of these algorithms are also used in a non-supervised context:
\begin{itemize}
	
	\item \textbf{MLPR} --- This is similar to the MLPC, but performing a regression
	task instead of a binary classification. In this case the number of neurons per \
	layer is allowed to go up to 5000.
	 
	\item \textbf{LSTMR} --- The model architecture is identical to the LSTMC,
	but it is trained to perform regression.
	
\end{itemize}
Gradient-boosted decision trees are also well-suited to perform regression tasks, but we prefer to focus on MLP that features similar performances with much more efficient training (see Fig.~\ref{fig:F2:best} and Table~\ref{tab:trainingtime}).

\subsection{Training process}
\label{sec:training}

The synthetic dataset contains series of 174,720 time steps, namely 20 years of 364 days (exactly 52 weeks of 7 days for periodicity), and 24 hours per day.
A random sample forming 20~\% of the data or 34,944 time steps is used as test set.
For supervised algorithms, the remaining 80~\% is used for training.
For unsupervised algorithms, 20~\% of the training set is further reserved for optimizing the detection threshold.

The training is performed independently for all 33 selected generators on the 3 networks. 
Hyperparameters are optimized on a case-by-case basis using a 5-fold cross-validation, followed by a refit on the entire training set for the best choice of parameters.
For the MLP algorithms, the training is limited to 200 epochs and terminates earlier if convergence of the $F_2$ score has occurred. LSTM algorithms are trained on a maximum of 50 epochs as they are more expensive.

\newpage

\section{Results}
\label{sec:results}

In this section, we present the results of a systematic investigation of all input parameters, algorithms and hyperparameters, for the three considered power grids.

\subsection{Hyperparameters}
\label{sec:hyperparameters}

The optimal hyperparameters generally fall inside the selected range, hence validating the choice of the range. The RFC requires 80 trees on average. The GBC often gives optimal results with 10 reinforcement stages, even though this number climbs up to the maximum value 1000 in a few cases. The MLPR does not use more than 1000 neurons per layer, and the final number of layers for MLP algorithms is between 2 and 4.

Two notable exceptions are found. First, the KNNC algorithm gives best results with $k = 1$. Combined with the poor overall performance reported below, this indicates that it lacks the capability of generalization required to address the difficult classification problem

The second exception regards the LSTM algorithms that perform best with the maximal number of stacked layers (3). This suggests that better performances could be achieved with a larger number of layers. The limitation in our case is computing power.

\begin{figure*}
	\begin{subfigure}{0.33\linewidth}
		\includegraphics[width=\linewidth]{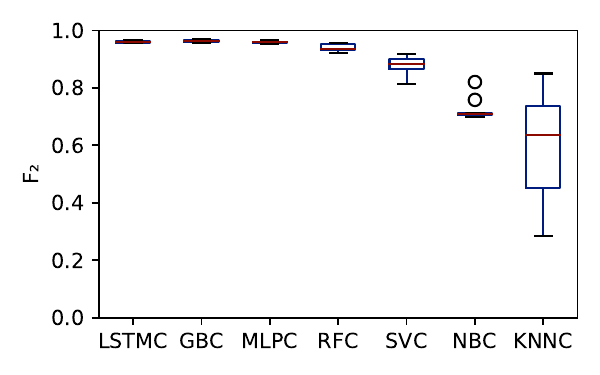}
		\caption{Switzerland}
		\label{fig:F2:classifiers:CH}
	\end{subfigure}
	\begin{subfigure}{0.33\linewidth}
		\includegraphics[width=\linewidth]{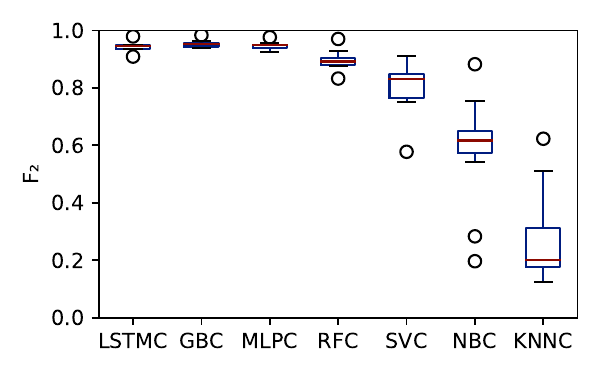}
		\caption{Spain}
		\label{fig:F2:classifiers:ES}
	\end{subfigure}
	\begin{subfigure}{0.33\linewidth}
		\includegraphics[width=\linewidth]{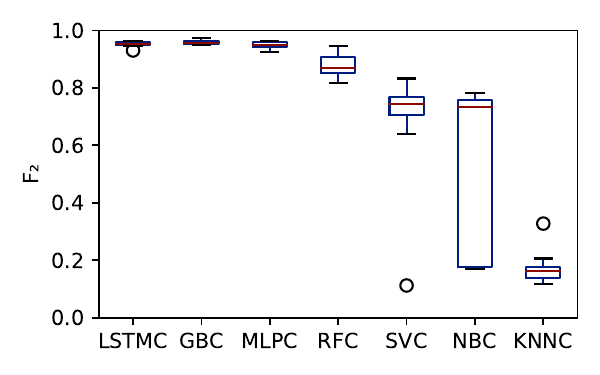}
		\caption{Germany}
		\label{fig:F2:classifiers:DE}
	\end{subfigure}
	\caption{$F_2$ score for the 7 classifier algorithms on the test set.
	In each case the median value over all the selected power plants
	is represented by the red line,
	surrounded by a box representing the first and third quartile,
	and whiskers at the minimum and maximum.
	Isolated outliers are denoted with circles.}
	\label{fig:F2:classifiers}
\end{figure*}
\begin{figure*}
	\begin{subfigure}{0.33\linewidth}
		\includegraphics[width=\linewidth]{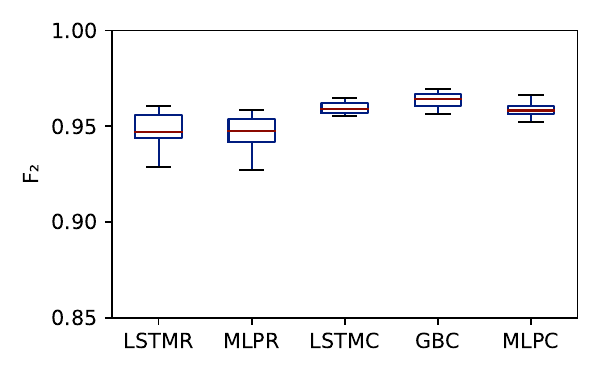}
		\caption{Switzerland}
		\label{fig:F2:best:CH}
	\end{subfigure}
	\begin{subfigure}{0.33\linewidth}
		\includegraphics[width=\linewidth]{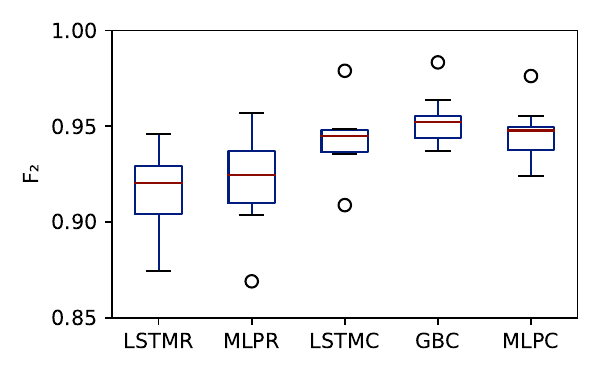}
		\caption{Spain}
		\label{fig:F2:best:ES}
	\end{subfigure}
	\begin{subfigure}{0.33\linewidth}
		\includegraphics[width=\linewidth]{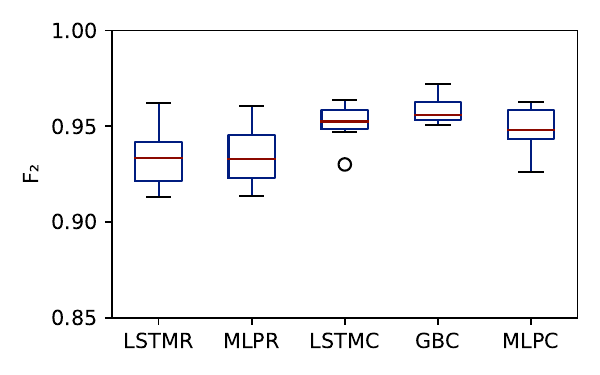}
		\caption{Germany}
		\label{fig:F2:best:DE}
	\end{subfigure}
	\caption{$F_2$ score for the 5 best algorithms on the test set,
	including the two unsupervised algorithms.
	Boxes and whiskers are as in Fig.~\ref{fig:F2:classifiers}.}
	\label{fig:F2:best}
\end{figure*}

\subsection{Comparison between algorithms}
\label{sec:algo-comparison}

The first important difference between algorithms is the time they require for training, as this varies over three orders of magnitude. 
Table~\ref{tab:trainingtime} shows the difference for a benchmark situation with fixed input vectors of size $180 \times 139,776$: this is the Swiss network with generation only (36 units) and a 4-time-step history including the entire context ($4 \times 36 = 144$), over 16 years of training data.
The training time varies with the size of the input vector, although not linearly and differently for all algorithms. LSTM algorithms are trained on different machines with large GPUs, so they cannot be directly compared with the others, but it is clear that they require more operations and therefore a longer training time on a comparable hardware.
We do not observe a significant difference in training times between the supervised and unsupervised versions of the MLP and LSTM algorithms.

\begin{figure}
	\centering
	\includegraphics[width=\linewidth]{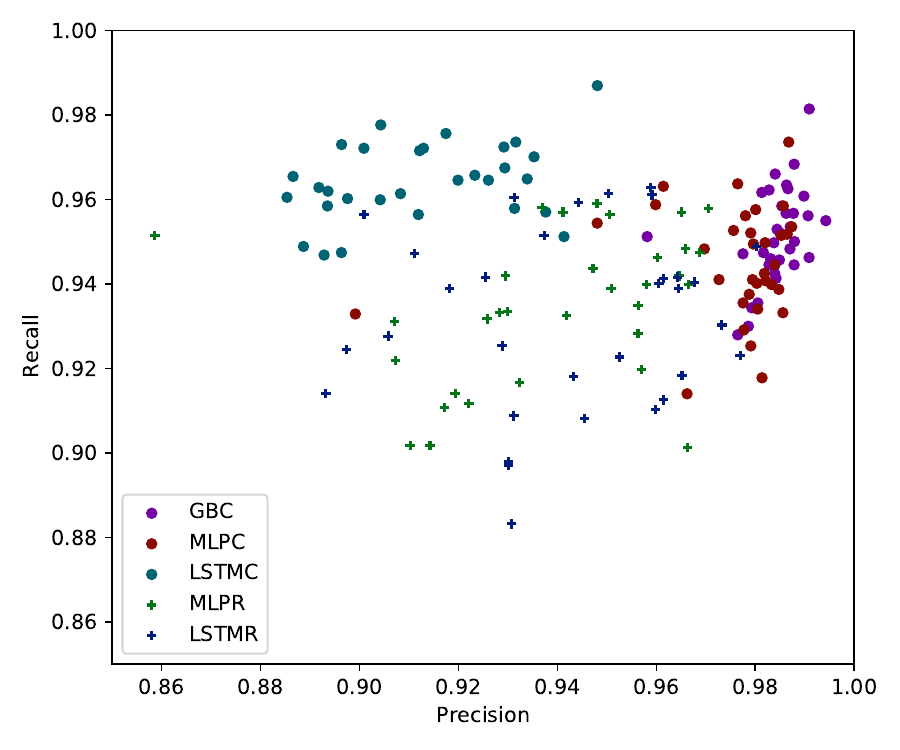}
	\caption{Precision (one minus rate of false positives) and recall (false negatives)
	for the five best performing algorithms on all 33 selected power plants
	of the three national grids, in each case with the best choice
	of hyperparameters. The circle/crosses correspond respectively
	to supervised/unsupervised algorithms.}
	\label{fig:precision-recall}
\end{figure}

The performance of the 7 classifier algorithms on the 3 different networks is measured by the $F_2$ score~\eqref{eq:F2} on the test set, and reported in Figure~\ref{fig:F2:classifiers}.
Three algorithms perform best: GBC, LSTMC, and MLPC.
The difference between them is not significant in terms of $F_2$ score, but they do actually perform differently in terms of precision and recall, as shown in Figure~\ref{fig:precision-recall}. The deep neural network LSTMC misses few actual attacks, with a rate of false negatives always below 5~\%, but it raises false positive alerts in up to 12~\% of the cases. On the contrary, the MLPC and GBC have a higher precision, with typically less than 5~\% false positives, but they do not achieve the same detection capability as the LSTMC. 
Since the training capability of the LSTMC is limited by the available computing power, it is quite possible that the precision of the LSTMC algorithm could be further improved with longer training times and larger architectures.

The good performance of the unsupervised algorithms is immediately visible in Fig.~\ref{fig:series}, as the prediction (dashed line) follows quite accurately the ground truth (solid line). The coefficient $R^2$ measured on the validation set is typically above 0.9, averaging at 0.95 and reaching 0.97 in the best cases, for all power plants on all 3 networks, both for the MLPR and LSTMR.
The relative error is defined as the root mean square error of the prediction divided by the power plant's rated power. It is typically of the order of 6-7~\% and varies between 2~\% and 16~\% depending on the power plant. The relative error is similar for both algorithms, even though the LSTMR outperforms the MLPR by 1~\% on average.
When applied to the classification of cyber-attacks, this leads to $F_2$ scores that are nearly as good as the supervised algorithms, as shown in Fig.~\ref{fig:F2:best}.
They are however much more versatile when it comes to detecting other type of anomalies.

\begin{table}
	\footnotesize
	\centering
	\begin{tabular}{|c|c|c|c|c|c|c|}
		\hline 
		\textbf{Algorithm} & NB & KNN & SVC & RFC & GBC & MLPC 
		\\
		\hline
		\textbf{Time [min]} & 0.05 & 11.6 & 0.26 & 2.8 & 85.9 & 7.33
		\\
		\hline
	\end{tabular}
	\caption{Training time in minutes for all classification algorithms
	except the LSTMC, on a standard laptop, for an input vector
	of size $180 \times 139,776$. The given time is an average 
	over the 10 selected hydroelectric power plants of the Swiss network,
	and it includes in each case hyperparameter optimization.}
	\label{tab:trainingtime}
\end{table}

\subsection{Comparison between types of input}
\label{sec:input-comparison}

\begin{figure}
	\includegraphics[width=\linewidth]{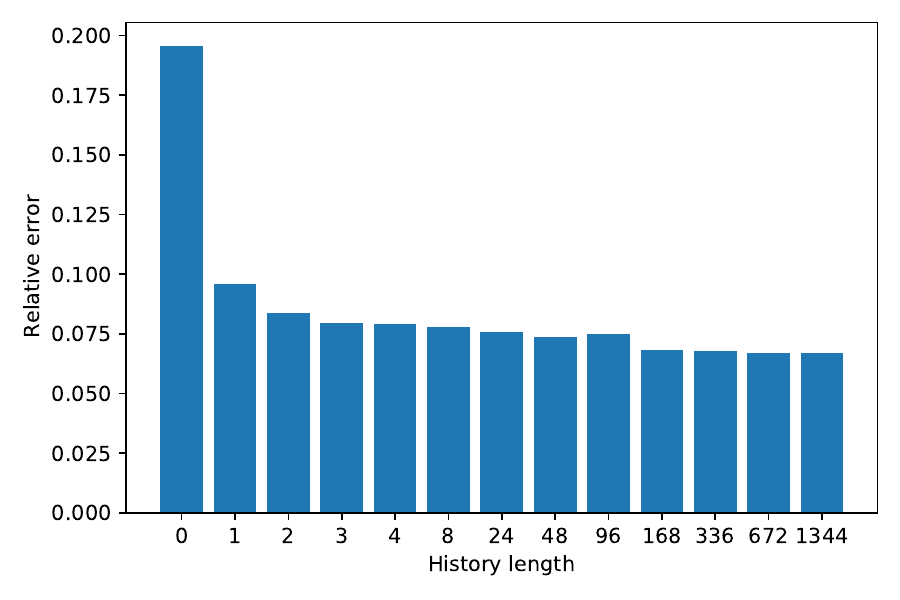}
	\caption{The relative error of the MLPR algorithm as a function of
	the length of the history used as input data.
	The reported error is an average over the 10 selected 
	generators in the Swiss network.}
	\label{fig:history}
\end{figure}

\begin{figure*}
	\centering
	\begin{subfigure}{0.48\linewidth}
		\includegraphics[width=\linewidth]{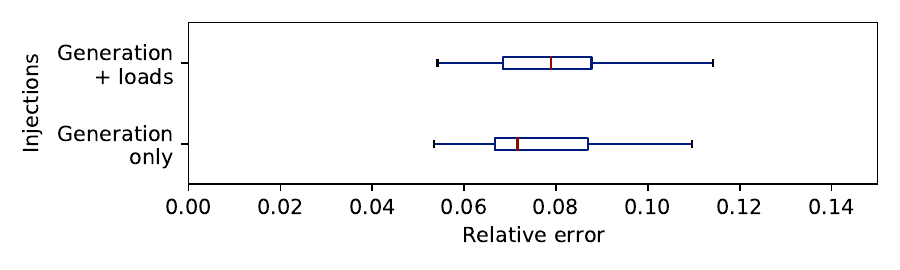}
		\caption{}
		\label{fig:inputs:injections}
	\end{subfigure}
	\begin{subfigure}{0.48\linewidth}
		\includegraphics[width=\linewidth]{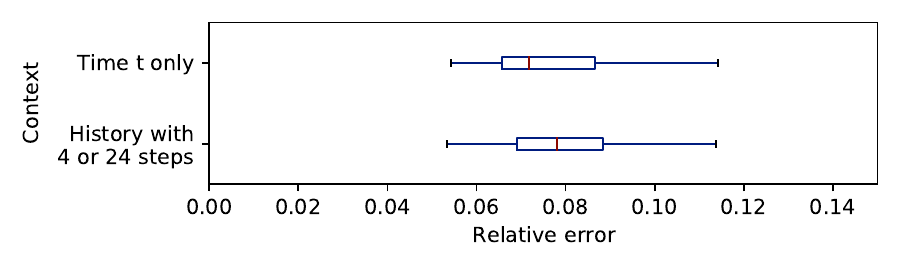}
		\caption{}
		\label{fig:inputs:context}
	\end{subfigure}
	\caption{Relative error of the unsupervised MLPR algorithm
	depending on the choice of input vectors,
	for the 10 selected generators in the Swiss network:
	(a) with or without the information on the network loads;
	(b) with the context at time $t$ only or including
	the context over the entire history (4 or 24 time steps).
	In all cases, only the best choice among all the marginalized
	parameters is taken into account.}
	\label{fig:inputs}
\end{figure*}

In order to report the results of experiments with different input parameters in the simplest way, we focus on the MLP algorithms that provide the best compromise between performance and training time. Moreover, we use the non-supervised MLPR algorithm as benchmark as the relative error provides a more representative metric than the $F_2$ score for generic anomalies.
We have nevertheless verified that similar results are found with all other algorithms.

Three important points emerge from our analysis.
The first is shown in Fig.~\ref{fig:inputs:injections}: the performance does not improve when the load measurements are included in the input data. 
This may be surprising, as the information about the loads appears to be relevant: both the total load that must be matched by the production sources and the distribution of loads in the network are playing a role in determining the generation dispatch.
We conjecture that this may be due to the dilution of the information among the entries of the input vectors, possibly making it more difficult for the algorithm to isolate the relevant signals.

The second point concerns the inclusion of historical data. Fig.~\ref{fig:history} clearly shows how the performance drastically improves as soon as history is taken into account, with the relative error of the MLPR algorithm dropping from around 20~\% to less than 10~\% already with the information of the previous time step. The performance   keeps improving as more time steps are included, but only in a very mild way.
We conclude that considering few historical time steps, typically less than 24, is sufficient to guarantee the maximal accuracy of our anomaly detection algorithm.

The last point is about the context: we compare in Fig.~\ref{fig:inputs:context} the performance of the MLPR algorithm with the context given at time $t$ only, and including the history of the context generators, tested in this case with a history length of 4 and 24 time steps. It should be noted that the size of the input vector increases drastically in this case: if the network contains $N$ generators, there are $N-1$ generators forming the context, and including 24 time steps of history for all of them amounts to considering a vector of size $25 N - 1$, whereas using the 24-step history of the target generator only requires an input vector of size $N + 23$.
In spite of the large amount of additional data available to the algorithm, Fig.~\ref{fig:inputs:context} shows that the performance does not improve when the history of the context is taken into account. In fact, the median value of the error is  even slightly lower without that additional data.

\subsection{Robustness against multiple attacks}
\label{sec:multiple-attacks}

Finally, we test the robustness of the anomaly detection algorithms against multiple, concurring attacks.
This is measured at inference time, after the algorithm have been trained without simultaneous attacks.
For each selected generator and its corresponding algorithm, we add on/off anomalies on other generators of the same network, effectively modifying the features corresponding to the context at time $t$.

We present here the results of the Swiss network, since the influence of context errors is expected to be larger in smaller networks.
We also restrict our attention to the MLPR algorithm with 24 hours of history, context at time $t$ only, and generator data only. This is one of the most interesting cases since the sensitivity to concurring attacks is more subtle in a non-supervised approach with minimal inputs.
We have checked that the results are similar in other situations.

\begin{figure*}
	\centering
	\begin{subfigure}{0.48\linewidth}
		\includegraphics[width=\linewidth]{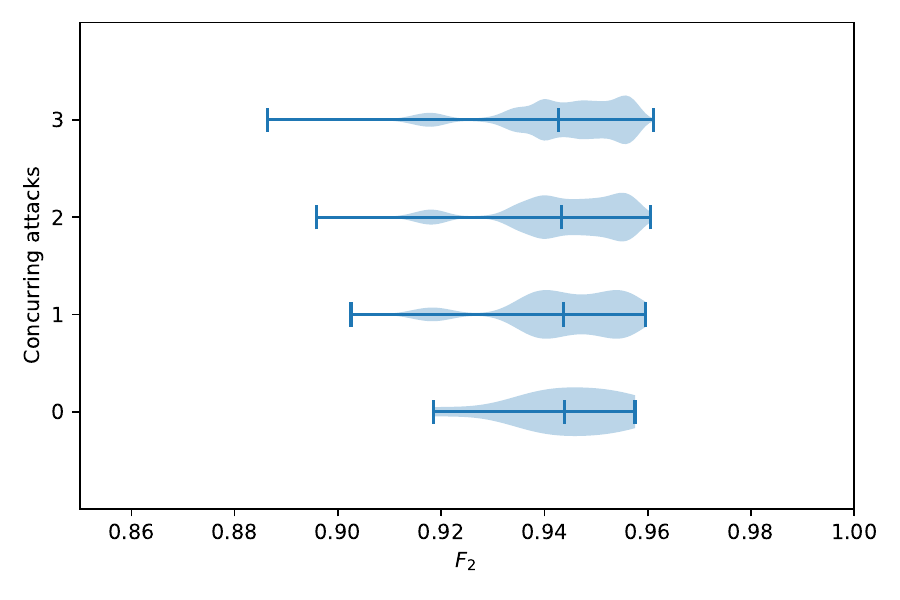}
		\caption{}
		\label{fig:multiple-attacks:F2}
	\end{subfigure}
	\begin{subfigure}{0.48\linewidth}
		\includegraphics[width=\linewidth]{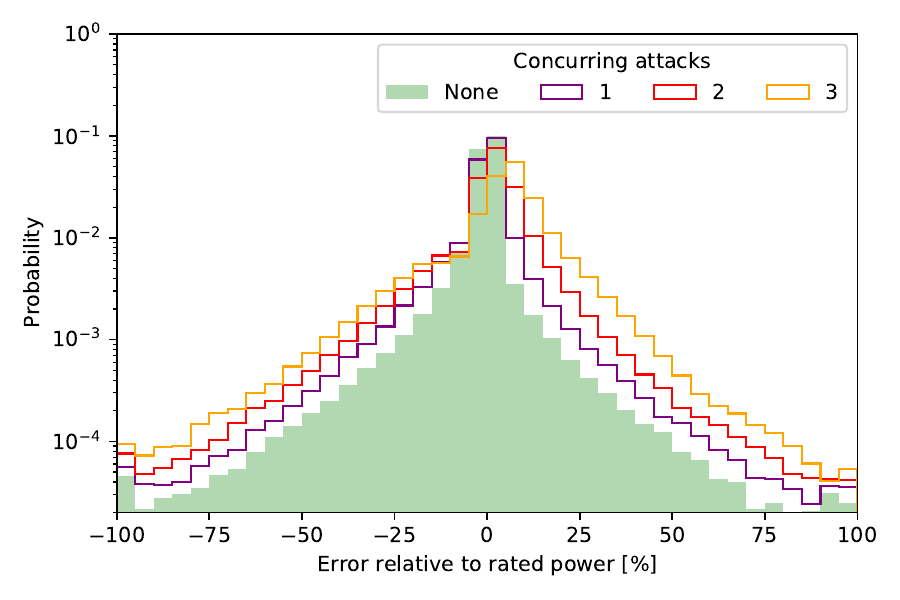}
		\caption{}
		\label{fig:multiple-attacks:errors}
	\end{subfigure}
	\caption{Performance of the MLPR algorithm against multiple,
	concurring attacks:
	(a) Distribution of $F_2$ scores for individual generators of the Swiss
	network, in the basic setup (single anomaly)
	as well as with 1, 2, or 3 concurring on/off attacks
	distributed on all combinations of other generators in the network;
	(b) Corresponding error relative to the rated power of the generators,
	when the most effective concurring attacks are selected.}
	\label{fig:multiple-attacks}
\end{figure*}

In this setup, there are 35 generators forming the context, and therefore respectively 35, 595, and 6545 combinations of 1, 2, and 3 concurring attacks.
The results of our analysis are given in Fig.~\ref{fig:multiple-attacks}.
On average, the $F_2$ score is barely affected by multiple, concurring attacks, as shown in Fig.~\ref{fig:multiple-attacks:F2}. This is because many concurring attacks have little effect on the MLPR prediction, and random combinations of attacks typically cancel out.
However, we see clearly that the tail of the distribution of $F_2$ scores is lowered in the presence of more attacks:
carefully selecting which generators to attack concurrently can significantly enhance the ability to hide each individual attack.
Nevertheless, in the cyber-security perspective, this requires a good knowledge of the operational state of the grid as well as the ability to perform attacks on chosen generators, both of which are rather unlikely.

Even when the most effective concurring attacks are chosen, the prediction of the MLPR algorithm remain relatively robust, as illustrated in Fig.~\ref{fig:multiple-attacks:errors}. In this case, we observe that the rate of sizable relative errors is multiplied by about 1.5 for each additional concurring attack, but this rate remains quite low (note the logarithmic scale).
The effect of a single concurring attack on the MLPR algorithm is also reported in Fig.~\ref{fig:series}, where it is shown as a band surrounding the prediction.

\section{Discussion and conclusions} 
\label{sec:discussion}

\begin{figure*}
	\begin{subfigure}{0.33\linewidth}
		\includegraphics[width=\linewidth]{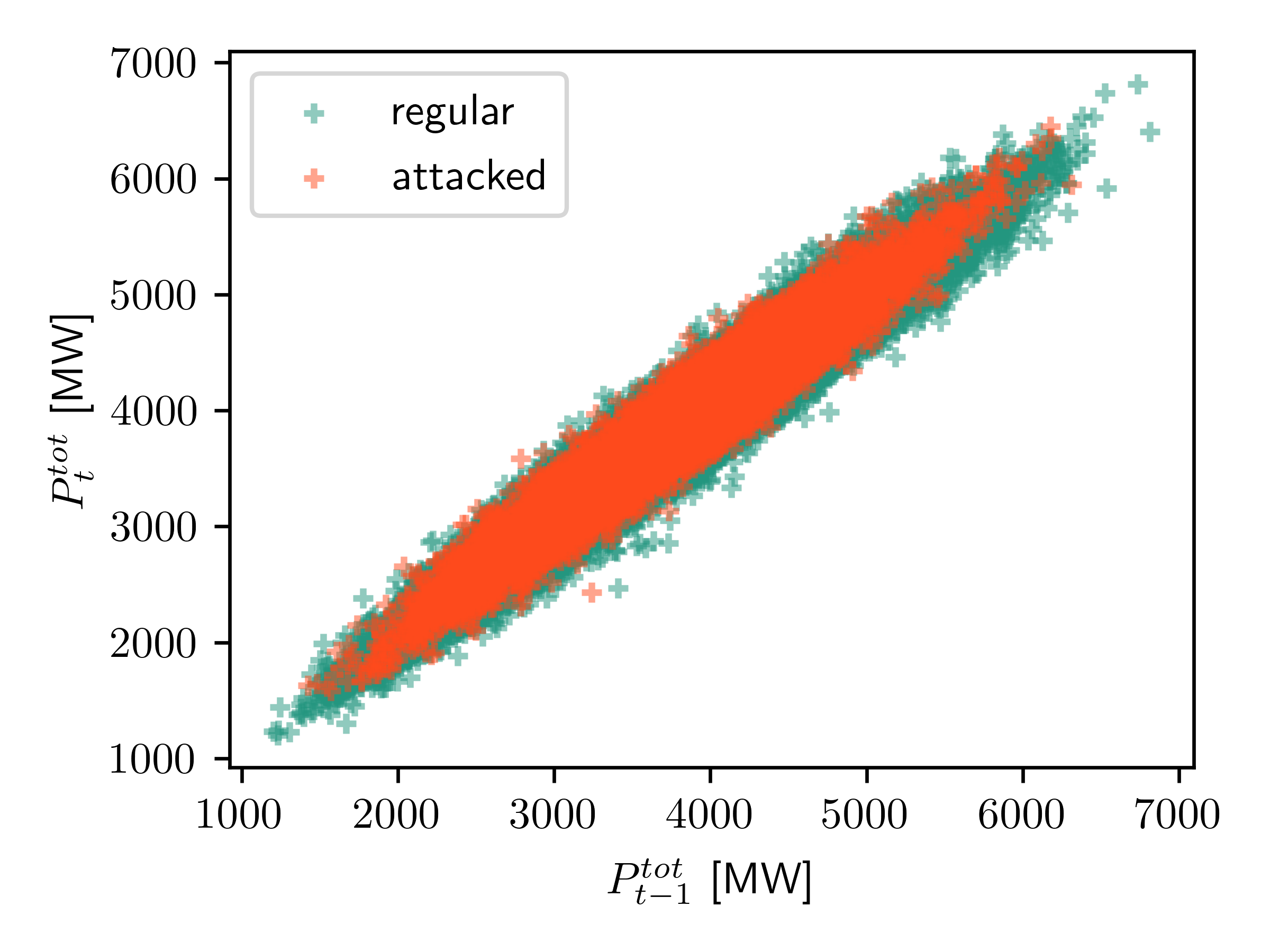}
		\caption{}
		\label{fig:anomalies:a}
	\end{subfigure}
	\begin{subfigure}{0.32\linewidth}
		\includegraphics[width=\linewidth]{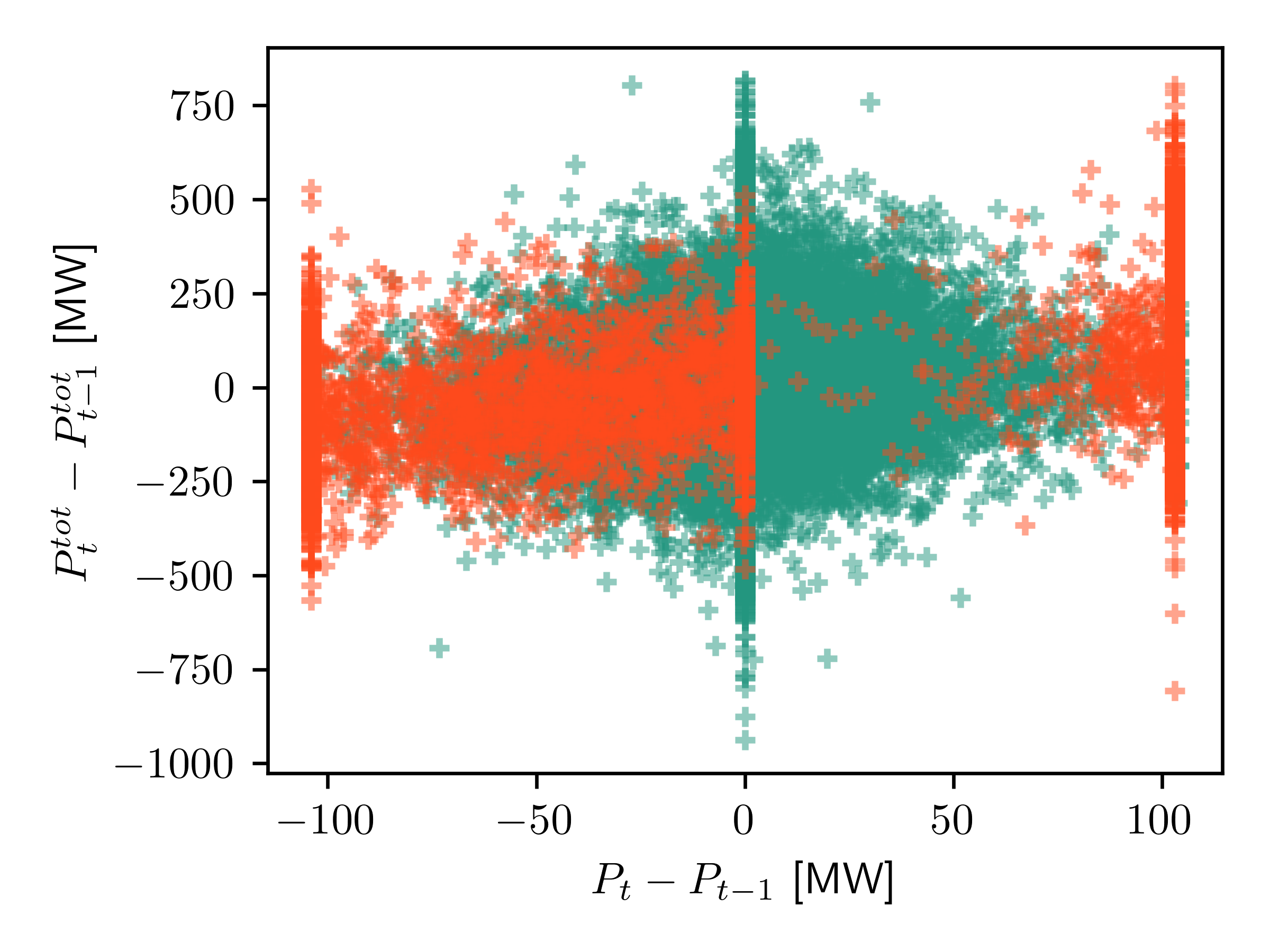}
		\caption{}
		\label{fig:anomalies:b}
	\end{subfigure}
	\begin{subfigure}{0.32\linewidth}
		\includegraphics[width=\linewidth]{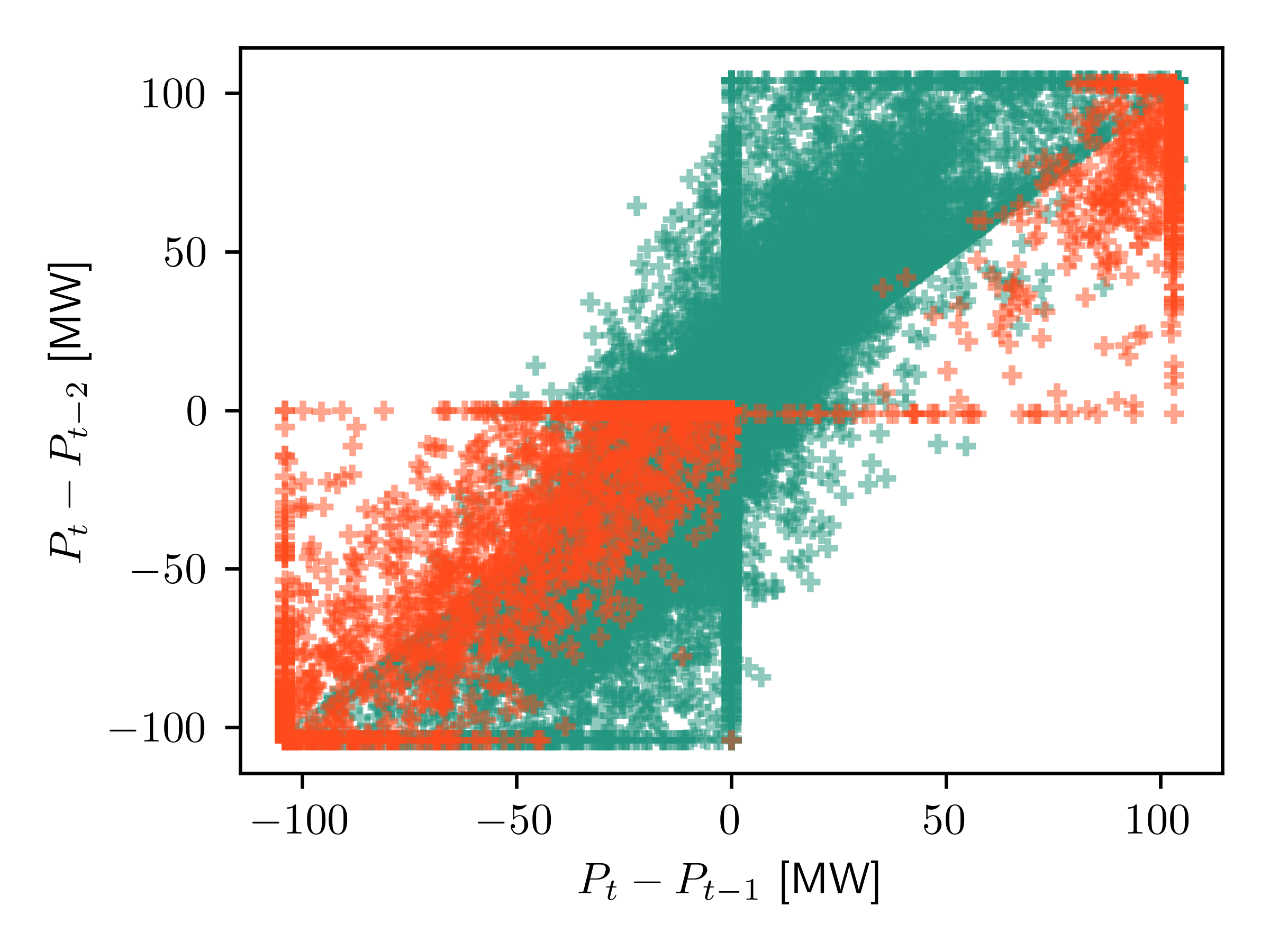}
		\caption{}
		\label{fig:anomalies:c}
	\end{subfigure}
	\caption{Distribution of anomalies compared to the regular situation
	shown with different metrics.
	Panel (a) shows the correlation between the total production in the Swiss network
	at consecutive time steps $t-1$ and $t$.
	Panel (b) shows the variation of the total production against the 
	variation of a chosen generator (Cavergno).
	The denser regions at the extremities of the horizontal axis contain both regular
	and attacked points, even though the former are hidden.
	Panel (c) shows the variation over two consecutive time steps against
	the variation over one time step for the same chosen generator.}
	\label{fig:anomalies}
\end{figure*}

The results presented in the previous section are characterized by a stark contrast between the good performance of the neural networks (MLPC, GBC, LSTMC, MLPR, LSTMR), and the significantly lower performance of traditional algorithms.
In particular, the k-nearest neighbors classifier (KNNC) performs very poorly, even worse than the baseline naive Bayes method (NBC). Even the support vector machines classifiers (SVC) that have been successfully used in the literature \citep{Esmalifalak17} do not perform very well on our dataset.
This difference can be explained by the fact that anomalies are purely contextual: the anomalous situations that we consider are well inside the space of all possible grid injections, at any given time $t$ but also as a distribution in time. They are only remarkable once the context is taken into account. This is illustrated in Fig.~\ref{fig:anomalies}, where we show the distribution of regular and anomalous situations along several metrics. Figs.~\ref{fig:anomalies:a} and \ref{fig:anomalies:b} show that the two distributions are mostly superimposed, even if they differ obviously in many aspects.
On the contrary, Fig.~\ref{fig:anomalies:c} shows that some patterns are only found in anomalous data, such as on-off-on and off-on-off sequences that are very rare in normal operating conditions.
Nevertheless, only a fraction of the anomalies can be isolated in this way.
The majority of anomalies are purely contextual, and hence inherently difficult to detect.
From this perspective, the overall good performance of the neural networks is very satisfactory. It is in fact quite remarkable and satisfying that the non-supervised algorithms perform nearly as well as the best classifiers, even though their regression task is significantly more involved.

Another interesting aspect of our results is the comparison between the different networks under consideration. The first observation is that the results are quite comparable on all three networks, despite their important topological differences.
However, there are sizable differences visible in Fig.~\ref{fig:F2:best}. 
The best performances are obtained on the Swiss network. This is not surprising, as it is the smallest network, and also the most homogeneous in terms of production type: besides a few nuclear power plant with flat production profiles, only hydro generators are connected to the Swiss transmission grid. The behavior of the Swiss production is presumably more predictable as a whole, because all elements obey similar rules. The situation is quite different in Germany and Spain, where many different types of production are mixed. In fact, all neural networks perform nearly as well on the German grid than they do on the much smaller Swiss grid, whereas the performance is more variable in Spain, with very good results on some power plants, and less so on others. We were not able to pinpoint a single factor explaining the discrepancy in performance within the Spanish network, but a possible explanation has to do with the proximity of sources in terms of network distance.
The Spanish network is by far the largest, with nearly twice as many nodes as the German one, but it has significantly fewer generators, hence a distribution that is much more sparse, see Fig.~\ref{fig:networks}.

The analysis of various input vectors also carries an important message.
Even though taking both context and history into account is crucial for the detection of contextual anomalies, there is no need to include large amounts of input data to obtain good results.
A reasonable trade-off between detection performance and usage of resources consists in taking 24 time steps of history for the considered generator, namely an entire day, and a context made of other generators at time $t$ only. For a network with $N$ generators, this requires only $N + 23$ entries. With this choice, the shallow neural networks of the MLPC and MLPR algorithms can be trained very quickly.

Further improvement of the $F_2$ scores could possibly be achieved with deeper neural network architectures, such as enhanced version of the LSTMC and LSTMR algorithms,  possibly with longer histories. This comes at a high computing cost, however, and should include real time series as input.
This perspective is left for industry-grade applications.
We note that combining different algorithm is in an interesting possibility in this regard. While the LSTM and MLP algorithm have similar $F_2$ scores, the distribution of errors is actually quite different  in either case, as illustrated in Fig.~\ref{fig:precision-recall}.
This means that their combination cannot be used to improve the $F_2$ score significantly, but it can be used to tune the balance between precision and recall, so as to lower the rate of false positives, for instance.

\section*{Acknowledgment}

This work has been supported by the Cyber-Defence Campus of armasuisse.

\newpage

\bibliographystyle{elsarticle-harv}
\bibliography{bibliography}

\end{document}